\documentstyle[mncite,psfig]{mn2e}
\title[Highly extinguished outflows in PKS 1345+12]{Highly extinguished
emission line outflows in the young radio source PKS
1345+12}
\author[J. Holt et al.]{J. Holt$^{1}$\thanks{E-mail:
j.holt@sheffield.ac.uk}, C. N. Tadhunter$^{1}$ and R. Morganti$^{2}$\\
$^{1}$Department of Physics and Astronomy, University of Sheffield,
Sheffield,  S3 7RH, UK.\\
$^{2}$Netherlands Foundation for Research in Astronomy, Postbus 2,
7990 AA Dwingeloo, The Netherlands}

\newcommand{\msun}{M$_{\odot}$}
\newcommand{\oxythr}{{[O III]}}

\newcommand{\oxyonew}{{[O I]}$\lambda\lambda$6300,6363}
\newcommand{\sulphurtwo}{{[S II]}}

\begin{document}
\maketitle
\begin{abstract}
We present new, intermediate resolution spectra ($\sim$ 4 \AA) of the
compact radio source PKS 1345+12 (4C 12.50, z = 0.122) with large spectral
coverage ($\sim$ 4500 \AA). Our spectra clearly show extended line emission 
up to $\sim$ 20 kpc from the nucleus. This is
consistent with the asymmetric halo of diffuse emission observed in
optical and infra-red images. At the position of the nucleus we
observe complex emission line profiles. Gaussian fits to the {[O III]}
emission lines require 3 components (narrow, intermediate and broad),
the broadest of which has FWHM $\sim$ 2000 km s$^{-1}$ and is
blue shifted by up to $\sim$ 2000 km s$^{-1}$ with respect to the halo
of the galaxy and HI absorption. We interpret this as material in
outflow. We find evidence for high reddening and measure E(B-V) $>$ 0.92
for the broadest, most kinematically disturbed component. This
corresponds to an actual H$\beta$ flux 130 times brighter than that observed. 
From our model for {[S II]}$\lambda\lambda$6716,6731 we estimate
electron densities of n$_{e}$ $<$150 cm$^{-3}$, n$_{e}$ $>$5300
cm$^{-3}$ and  n$_{e}$ $>$4200 cm$^{-3}$ for the regions emitting the narrow,
intermediate and broad components respectively. We calculate a total
mass of line emitting gas of M$_{gas}$ $<$ 10$^6$ M$_{\odot}$.
Not all emission line profiles can be reproduced by the same
model with {[O I]}$\lambda\lambda$6300,6363 and {[S
II]}$\lambda\lambda$6716,6731 requiring separate, unique models.
We argue that PKS 1345+12 is a
young radio source whose nuclear regions are enshrouded in a dense
cocoon of gas and dust. The radio jets are expanding through this
cocoon, sweeping material out of the nuclear regions. Emission
originates from three kinematically distinct regions though gradients
(e.g. in density, ionisation potential, acceleration etc) must exist
across the regions responsible for the emission of the intermediate
and broad components. 
\end{abstract}


\section{Introduction}
\begin{center}
\begin{table*}
\begin{minipage}{146mm}
\caption{Log of observations. The red and blue arms were exposed 
simultaneously using a dichroic at 6100 \AA. }
\label{Table 1}
\begin{tabular}{lcllcccl}\\  \hline
\multicolumn{1}{c}{Date}&\multicolumn{1}{c}{Arm} & \multicolumn{1}{c}{Exposure}& \multicolumn{1}{c}{Setup} & \multicolumn{1}{c}{Slit PA} & \multicolumn{1}{c}{Slit width} & \multicolumn{1}{c}{Seeing} & \multicolumn{1}{c}{Conditions} \\
 & & \multicolumn{1}{c}{(s)} & \multicolumn{1}{c}{(CCD + grating)} & & \multicolumn{1}{c}{(arcsec)} & \multicolumn{1}{c}{(arcsec)} & \\ \hline
20010512 & R & 1*900 & TEK4 + R316R & 104 & 1.3 & 1.32 $\pm$ 0.16 & Photometric \\
20010512 & B & 1*900 & EEV12 + R300B & 104 & 1.3 & 1.32 $\pm$ 0.16 & Photometric\\
20010512 & R & 3*1200 & TEK4 + R316R & 160 & 1.3 & 1.32 $\pm$ 0.16 &Photometric\\
20010512 & B & 3*1200 & EEV12 + R300B & 160 & 1.3 & 1.32 $\pm$ 0.16&Photometric \\
20010514 & R & 3*1200 & TEK4 + R316R & 230 & 1.3 & 1.67 $\pm$ 0.12
 &Variable transparency\\
20010514 & B & 3*1200 & EEV12 + R300B & 230 & 1.3 & 1.67 $\pm$ 0.12
 &Variable transparency \\ \hline
\end{tabular}
\end{minipage}
\end{table*}
\end{center}

Gigahertz Peaked Spectrum radio sources (GPS: D $<$ 1 kpc) and the
more extended Compact Steep Spectrum radio sources (CSS: D $<$ 15 kpc)
make up a significant fraction ($\sim$ 40\%) of the radio source population, 
but their nature is not fully understood (see e.g. \pcite{odea98} and
references therein). Are these compact radio sources linked to the
large scale radio galaxies by some evolutionary sequence, or are they
indeed a separate class of object?

There are two main theories to describe compact radio sources: the
{\em frustration scenario} \cite{vanbreugel84}
and the {\em youth scenario} \cite{fanti95}. 
In the {\em frustration scenario}, the radio source is enshrouded by a
cocoon of gas and dust so dense that the radio jets cannot escape the
nuclear regions and the radio source is confined and frustrated for
its entire lifetime. Alternatively, the {\em youth
scenario} attributes compactness of the radio source
to evolutionary
stage -- if we observe a radio source when it
is relatively young, the radio jets will not have had the time to
expand much and will still be relatively small. In this scenario, the
GPS and CSS sources represent the earliest stages in the evolution of
radio galaxies, eventually expanding into the large scale doubles.
This is currently the preferred theory and is supported by
estimates of dynamical
ages of GPS sources 
of t$_{dyn}$ $\sim$ 10$^2$ - 10$^3$ years (e.g. \pcite{owsianik98};
\pcite{tschager00}) and radio spectral ages of the larger CSS sources
 of t$_{sp}$ $<$ 10$^4$ years
\cite{murgia99}. It has also been suggested that the host galaxies
pass through multiple, short-lived (t$_{rs}$ $<$ 10$^{4}$ years) activity
phases responsible for the extended, diffuse emission commonly found
around GPS sources (\pcite{readhead94}; \pcite{readhead96}), although
there are no strong arguments either for or against this theory \cite{odea98}.

There is much evidence to suggest that, in a significant fraction of
the radio source population, the radio sources are triggered by
merger events involving two or more galaxies, of which at least one is
gas rich \cite{heckman86}. Many compact radio sources exhibit features
attributed to mergers such as double nuclei, tidal tails, arcs of
emission and distorted isophotes, suggesting that 
these sources are possibly observed
relatively shortly after a merger event \cite{stanghellini93}, 
before the system has recovered and settled down. However, alternative
activity triggering mechanisms have been suggested such as
intracluster cooling flows \cite{bremer97}. Whilst mergers may be the
most likely trigger of the activity, it is clear that {\em all}
activity inducing processes involve the injection of substantial
amounts of gas and dust into the nuclear regions -- initially, at
least, the radio source will exist in a dense and dusty
environment. Subsequent evolution will be strongly dependent on both
the quantity and distribution of the circumnuclear material. If the
ISM forms an equatorial disk, the radio jets will undergo little
interaction with the ISM. On the other hand, if the ISM forms an
enshrouding cocoon, strong interactions are expected between the ISM
and the radio jets
\cite{bicknell97}, quasar-induced winds \cite{balsara93} or
any starburst-driven superwinds \cite{heckman90} in the early stages
of radio source evolution. Indeed, depending on the amount of material
in the cocoon, it may be necessary to hollow out bi-polar cavities in
the cocoon in order to allow the radio source to expand freely through
the halo of the host galaxy. Therefore, 
it seems likely that outflows of the ISM 
will be important in the early stages of radio source evolution.

Much of the work on compact radio sources has concentrated on the
 radio wavelength region, but
relatively little attention has been paid to other wavebands. High
quality optical spectra have the potential to 
provide important clues on the nature
of compact radio sources and their early evolution. However,
to date, most spectroscopic studies have
relied on poor quality, low resolution data
 (e.g. \pcite{gelderman94}). Recently, the 
promise of optical spectroscopic
 studies was demonstrated by observations of
 the compact, flat spectrum radio source PKS 1549-79 
 \cite{tadhunter01}. Low dispersion spectra of this radio source
revealed broad forbidden line profiles (FWHM $\sim$ 1350 km
 s$^{-1}$) blue-shifted by $\sim$ 600 km s$^{-1}$ with
 respect to the galaxy rest frame, believed to trace material in
 outflow. Combined with its compactness, \scite{tadhunter01} suggest
 PKS 1549-79 is a young radio galaxy whose radio jets, aligned close
 to the line of sight,  are sweeping material out of the nuclear regions.

\begin{center}
\begin{figure}
\centerline{\psfig{file=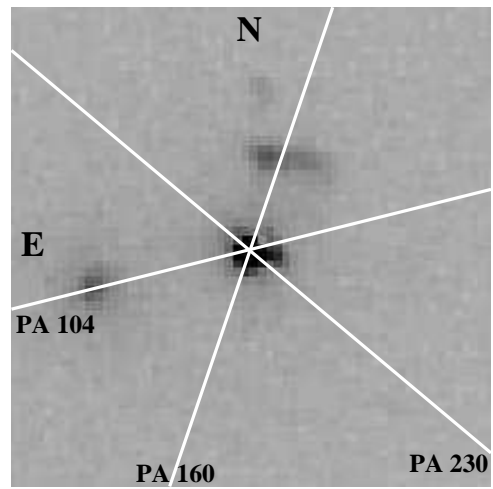,height=6.5cm,angle=0.}}
\caption{
Slit positions for the new spectroscopic observations of PKS
1345+12 superimposed on an O III
image taken with HST WFPC2 from 
\protect\scite{axon00}. Radio emission is
confined entirely within the western nucleus (at the centre of the image) on
a scale of $<$ 0.1 arcsec ($\sim$ 240 pc; 
\protect\pcite{stanghellini93}, \protect\pcite{stanghellini97}).
The second nucleus lies $\sim$ 1.8 arcsec (4.3 kpc) to the east and is
included in our spectra taken at PA 104. 
}
\label{Figure 1.}
\end{figure}
\end{center}

In this paper we present
new, high quality observations of the GPS source PKS 1345+12,
taken with ISIS, the dual-arm
spectrograph on the 4.2m William Herschel Telescope (WHT) on La
Palma. The high quality of the data for this source allows us to
investigate the putative circum-nuclear outflows in greater depth than
was possible for PKS 1549-79. The preliminary 
results of this study were reported in \scite{holt02}. 
ISIS is ideal for studies of this type as it allows spectra to be taken
with intermediate resolution without compromising the spectral range. 
By using a dichroic, the red and blue regions of
the spectrum were imaged simultaneously with CCDs optimised for each range.
We investigate the kinematics of the emission line gas in the nucleus
with respect to the galaxy halo to search for outflows -- a possible
impact of the nuclear activity on the ambient ISM.
We also investigate the physical properties
in the nuclear regions (e.g. density, gas mass) and outline a model for
compact radio sources using PKS 1345+12 as an example of a 
radio source in the early stages of radio source evolution.

\section{Previous observations of PKS 1345+12}
\begin{center}
\begin{figure*}
\centerline{\psfig{file=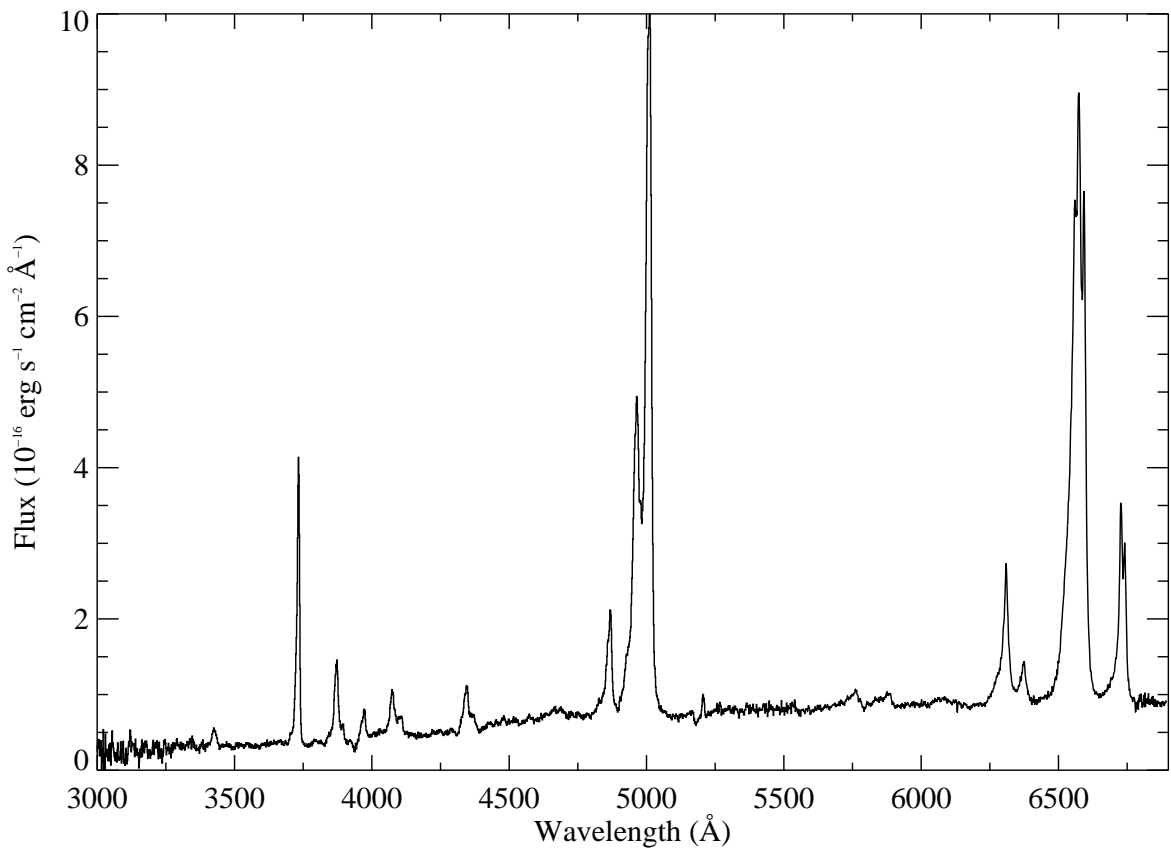,width=17cm,angle=0.}}
\caption{Composite spectrum of the nucleus (from PA 160; see sections
  3 and 4) at rest of PKS 1345+12. 
}
\label{Figure 2.}
\end{figure*}
\end{center}

The bright, compact radio source PKS 1345+12 is optically identified
with a 17th magnitude elliptical galaxy (e.g. \pcite{clarke66};
\pcite{burbidge70}; \pcite{gilmore86}). PKS 1345+12 has two nuclei
separated by $\sim$ 1.8 arcsec or 4.3 kpc\footnote{H$_{0}$ = 75 km
  s$^{-1}$, $q_0$ = 0.0 assumed throughout,
resulting in a scale of 2.37 kpc arcsec$^{-1}$ at $z$ = 0.122.}.
The western nucleus has a Seyfert 2/NLRG type spectrum
\cite{gilmore86}. The optical identification of the radio source is
controversial but the latest astrometry by \scite{stanghellini93}
associates the radio source with the western, highly reddened nucleus
(\pcite{evans99}, \pcite{smith89}, \pcite{stanghellini93}). 

Both nuclei are surrounded by an extended,
asymmetric halo detected in optical (\pcite{heckman86};
\pcite{gilmore86}) and near infra-red \cite{evans99} images.
There is clear evidence for
extended, distorted morphology including a strongly curved tidal tail
terminating $\sim$ 25 arcsec ($\sim$ 59 kpc) NW of the nucleus
(\pcite{heckman86}, \pcite{smith89}). \scite{gilmore86} estimate a 
projected linear size
of 110 $h^{-1}$ kpc (where $h$ is the Hubble constant with units of
100 km s$^{1}$ Mpc$^{-1}$).

There is much evidence to suggest that the nuclear regions of PKS
1345+12 contain large amounts of gas and dust. Deep, multiple
component HI 21cm absorption over a velocity range $>$ 700 km
s$^{-1}$ with column density N$_{\rmn HI}$/T$_{spin}$ = 6.1 $\times$
10$^{18}$ atoms cm$^{-2}$ 
has been detected by \scite{mirabel89} and more recently over
a velocity range $>$ 1000 km s$^{-1}$ by
\scite{morganti02}. The same velocity components are also
detected in CO (1$\rightarrow$0) emission (\pcite{mirabel89b},
\pcite{evans99}) and the shared CO halo indicates the galaxy is rich
in highly concentrated dust and molecular gas
(3.3 $\times$ 10$^{10}$ M$_{\odot}$; $\rho_{gas}$ $>$ 
2000 M$_{\odot}$ pc$^{-3}$; \pcite{evans99}). 

Both the distorted optical morphology and the presence of a rich ISM
suggest that PKS 1345+12 has been involved in a merger of two
galaxies in its recent history (\pcite{gilmore86}, \pcite{heckman86}). 
At least one of the galaxies must have been gas rich
(\pcite{stanghellini93}, \pcite{gilmore86}).
Indeed, the close proximity of
the two nuclei may indicate that the merger event began so recently
that it may still be ongoing (\pcite{gilmore86},
\pcite{heckman86}, \pcite{hutchings87}, \pcite{baum88},
\pcite{evans99}). PKS 1345+12 also has a significant young stellar
population (Robinson et al. 2003, in preparation) which may be the result of
a merger induced starburst. Further evidence for the prodigious star
formation in PKS 1345+12 is provided by the detection of a substantial
far-IR excess (L$_{\rm IR}$ = 1.7 $\times$ 10$^{12}$ L$_{\odot}$,
\pcite{evans99}) which leads to the classification of this source as
an ultra luminous infra-red galaxy (ULIRG).

The main radio source associated with PKS 1345+12 
is confined to a region $<$ 0.15 arcsec ($\sim$ 350
 pc) in size coincident with the western nucleus
 (\pcite{stanghellini93}, \pcite{stanghellini97}).
PKS 1345+12 has a typical GPS radio spectrum, peaking at $\sim$ 0.6
GHz with spectral index\footnote{S $\propto$ $\nu^{-\alpha}$ assumed  
throughout.}  -0.7 and  +0.9 above and below the peak respectively
 \cite{stanghellini98} and a 
sharp spectral cut-off near 400 MHz \cite{lister02a}. 

PKS 1345+12 has distorted, `triple' 
radio  morphology (i.e. a core and two jets are visible) aligned along
 PA 160 with a jet extending 0.04 arcsec (95 pc) to the SE before bending
 and expanding into a diffuse lobe. Weak  radio emission is detected
up to 0.03 arcsec (70 pc) NW of the bright knot towards the northen 
limit of the jet, tentatively identified as the core
 (\pcite{stanghellini97}). \scite{stanghellini01} also report very weak,
 extended emission on arc-second scales at 1.4 GHz resembling the
 parsec scale morphology.
Such extended weak emission may indicate
 PKS 1345+12 has experienced other activity phases in the past
(e.g. Stanghellini et al. 2002, private communication).

The radio morphology of PKS 1345+12 resembles only a few GPS
sources (see samples listed in \pcite{odea98}), though this may 
be a redshift issue -- PKS 1345+12 is one of the closest GPS sources
(z = 0.122) whilst most are at much higher redshift (z $>$ 0.5). The
S-shaped jet morphology is only detected in 5-10 \% of  GPS sources
and is attributed to recent AGN outflows where the black hole spin is
still precessing (\pcite{stanghellini01}, \pcite{lister02b}).
\scite{lister02a} detect superluminal motion (1.2c) in the southern
jet. 

Consistent with its status as a powerful radio source (P$_{5 GHz}$ =
10$^{26}$ W Hz$^{-1}$; \pcite{odea98} and references therein),
the nucleus of PKS 1345+12 emits strong emission lines. However,
although classified as an NLRG by \scite{grandi77}, it has
very broad forbidden emission lines with complex
structures and strong blue
asymmetries.
\begin{enumerate}
\item  PKS 1345+12 has two distinct redshift
systems: one for the high ionisation lines (e.g. {[O III]}, 
{[Ne III]}), another redshifted by +0.0008
producing the hydrogen Balmer and low ionisation lines (e.g. {[O I]}, {[O
II]}, {[S II]} and {[N II]}). 
\item The emission lines are much broader
than in normal NLRGs and exhibit distinct blue wings with
the low ionisation lines (FWHM $\sim$ 1200 km s$^{-1}$) having  
different  line profiles to the high
ionisation forbidden lines and hydrogen Balmer lines (FWHM $\sim$ 1600 km
s$^{-1}$). 
\end{enumerate}

In addition to the high ionisation optical forbidden line emission,
further evidence that PKS 1345+12 contains a powerful quasar nucleus
is provided by the detection of a broad (FWHM $\sim$ 2600 km s$^{-1}$)
component to Pa$\alpha$ \cite{veilleux97}, the detection of a point source
component in high resolution near-IR images \cite{evans99}, and the 
detection of a nuclear X-ray source \cite{odea00}.

\section{New spectroscopic observations of PKS 1345+12}
\begin{figure*}
\begin{minipage}{150mm}
\centerline{\psfig{file=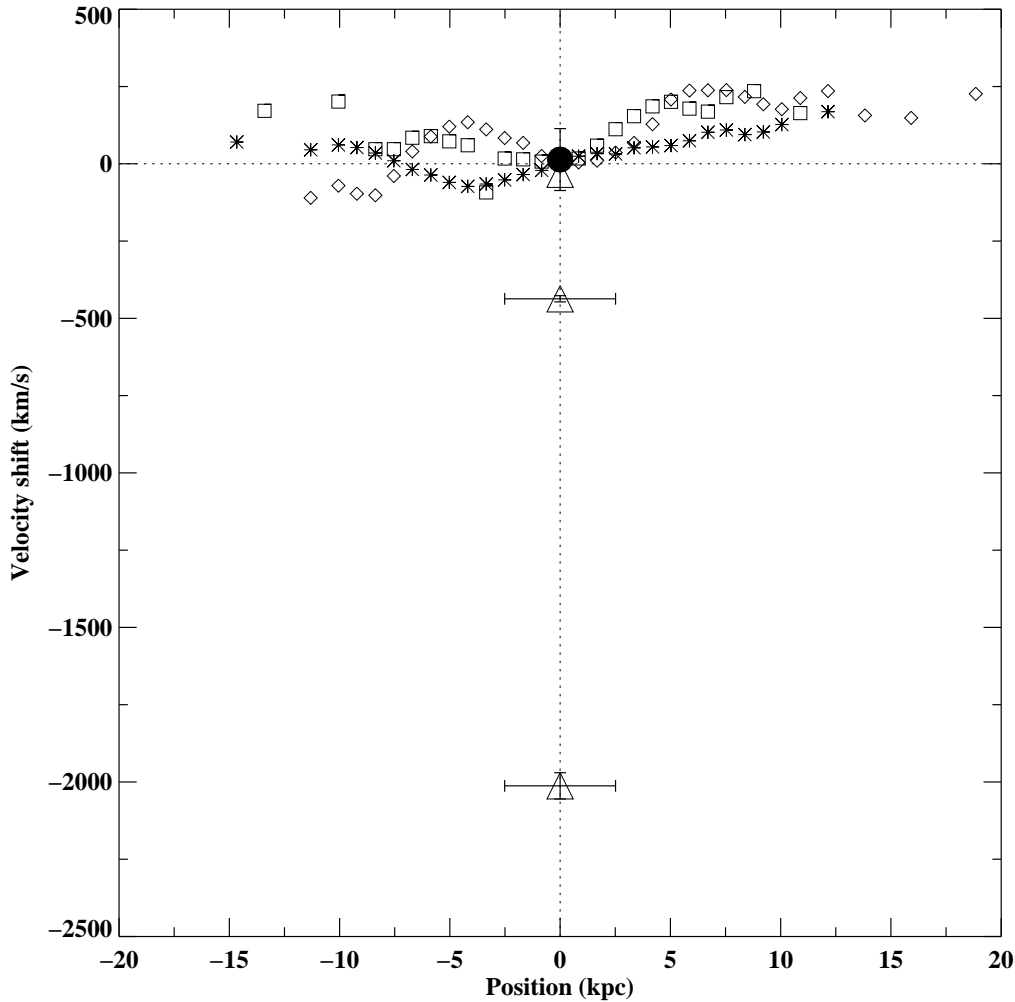,height=14cm}}
\caption{Spatial variations of the {[O II]} radial velocities
of the extended gaseous halo along all three PAs: PA 160  
(open diamonds), PA 104 (open squares) and PA 230 (asterisks). 
The open diamonds, squares and asterisks represent
the spatially extended component of {[O II]} (the error bars are small
and are incorporated in the points). Overplotted is the radial
velocity of the deep HI 21cm absorption (filled circle; \protect\pcite{mirabel89})
and the three components of {[O III]} in the nucleus (open triangles;
see section 4.2). Note, the orientations of the three PAs are as
 follows: i) PA 160: positive kpc = south, negative kpc = north; ii)
 PA 104: positive kpc = east, negative kpc = west; iii) PA 230:
 positive kpc = south-west, negative kpc = north-east. The vertical dotted line represents
  the centroid of the nuclear aperture. The horizontal dotted line
  represents the velocity at the position of the centroid.}
\label{Figure 3.}
\end{minipage}
\end{figure*}

We present new observations taken in May 2001 with ISIS, the dual-arm 
spectrograph using the 4.2m William Herschel Telescope (WHT) on La Palma.
In the red, the R316R grating was used with the TEK4 CCD, 
and in the blue, the R300B grating was used with the EEV12 CCD. A dichroic
at 6100\AA~was used to obtain spectra with wavelength range
6212-7720\AA~in the red and 3275-6813\AA~in the blue 
simultaneously. A summary of observations is presented in Table 1.
To reduce the effects of differential refraction, 
all exposures were taken when PKS 1345+12 was
at low airmass (sec $z$ $<$ 0.1) or along the parallactic angle.

Spectra were taken along three slit positions - PA 104, PA 160 and PA
230 (see Figure 1) with a 1.3 arcsec slit. 
PA 104 includes the second nucleus 1.8 arcesc (4.3
kpc) to the east of
the active nucleus. PA 160 lies along the radio axis and cuts through
the arc of emission 1.2 arcsec (2.8 kpc) to 
the north of the
nucleus. The final position, PA 230, was the parallactic angle of the
source at the time of observation. All angular scales are taken from 
\scite{axon00}.

The data were reduced in the usual way (bias subtraction, flat
fielding, cosmic ray removal, wavelength calibration, flux
calibration) using the standard
packages in IRAF. The two-dimensional spectra were also corrected for
spatial distortions of the CCD.
To reduce wavelength calibration errors due to
flexure of the telescope and instrument, separate arcs were taken at
each position on the sky. The final wavelength calibration accuracy,
calculated using the standard error on the mean deviation of
the night sky emission lines from published values
(\pcite{osterbrock96}) 
was 0.059\AA~and 0.112\AA~in the
red and blue respectively (PA 160). The spectral
resolution, calculated using the widths of the night sky emission lines
was 3.66 $\pm$ 0.09\AA~in the red and 4.54 $\pm$ 0.10
\AA~in the blue.

Comparison of several spectrophotometric standard
stars taken with a wide slit (5 arcsec) throughout the run gave a
relative flux calibration accurate to $\pm$ 5 per cent. This accuracy
was confirmed by good matching in flux between the red and blue spectra.
Further observations of standard stars with a narrow slit, 
matched to the slit width used to observe PKS 1345+12, were
used to correct for atmospheric absorption features (e.g. A and B
bands at $\sim$ 7600\AA~and $\sim$ 6800\AA~respectively). 

The main aperture used was the nuclear aperture -- 2.16 arcsec wide, centred
on the nuclear continuum emission

The spectra were extracted and analysed using the Starlink packages
FIGARO and DIPSO.

\vfill
\pagebreak

{\tiny
\begin{center}
\begin{table*}
\begin{minipage}{145mm}
\caption{Line data for all components of emission lines (where the
emission line is a doublet, data for the brightest line is tabulated) measured
  in the nucleus of PKS 1345+12 (see section 3). 
Columns are: $(a)$ emission line; $(b)$ emission line components where n, i and b
  are the narrow, intermediate and broad
components respectively; $(c)$ the quoted rest wavelength of the line
  in angstroms;
$(d)$ velocity shift of each component with
  respect to the narrow component for a particular
emission line (km s$^{-1}$); $(e)$ uncertainty on the velocity shift (km s$^{-1}$);
$(f)$ velocity width (FWHM, corrected for instrumental width) of the
emission line component  (km s$^{-1}$); $(g)$ uncertainty on the velocity
  width (km s$^{-1}$);
$(h)$ integrated line flux (10$^{-15}$ erg s$^{-1}$
cm$^{-2}$); $(i)$ error on the line flux (10$^{-15}$ erg s$^{-1}$).}
\label{Table 2}
\begin{tabular}{lccrrrrrr}\\  \hline
\multicolumn{1}{c}{Line}      & \multicolumn{1}{c}{Component} & \multicolumn{1}{c}{Rest}       & \multicolumn{1}{c}{Velocity}& \multicolumn{1}{c}{$\Delta$Velocity} & \multicolumn{1}{c}{Velocity} & \multicolumn{1}{c}{$\Delta$FWHM}& \multicolumn{1}{c}{Line} & \multicolumn{1}{c}{$\Delta$Flux} \\ 
          &           &\multicolumn{1}{c}{Wavelength} & \multicolumn{1}{c}{shift}    & \multicolumn{1}{c}{shift}  & \multicolumn{1}{c}{Width} & & \multicolumn{1}{c}{Flux} & \\
          &           & \multicolumn{1}{c}{(\AA)}    &\multicolumn{1}{c}{(km s$^{-1}$)} &\multicolumn{1}{c}{(km s$^{-1}$)} & (km s$^{-1}$) & (km s$^{-1}$)& \multicolumn{1}{c}{(10$^{-15}$)}& (10$^{-15}$) \\ 
\multicolumn{1}{c}{$a)$} & \multicolumn{1}{c}{$(b)$} & \multicolumn{1}{c}{$(c)$} & \multicolumn{1}{c}{$(d)$} & \multicolumn{1}{c}{$(e)$} & \multicolumn{1}{c}{$(f)$} & \multicolumn{1}{c}{$(g)$} & \multicolumn{1}{c}{$(h)$} &\multicolumn{1}{c}{$(i)$} \\
\hline
{[O III]} & n& 5006.9  &0       & 0   & 340 & 23 & 1.37 & 0.09 \\
          & i& 5006.9  &-402    & 9   & 1255& 12 & 19.78& 0.27 \\
          & b& 5006.9  &-1980   & 36  & 1944& 65 & 7.97 & 0.31 \\
{[O I]}   & n& 6300.3  &0       & 0   & 340 &    & 0.52 & 0.03 \\
          & i& 6300.3  &-48     & 9   & 1124& 33 & 2.32 & 0.10 \\
          & b& 6300.3  &-1095   & 66  & 2671& 91 & 2.37 & 0.13 \\
{[S II]}  & n& 6716.4  &0       & 0   & 340 &    & 1.28 & 0.03 \\
          & i& 6716.4  &-222    & 90  & 1124&119 & 1.42 & 0.34 \\
          & b& 6716.4  &-1374   & 213 & 2857&229 & 0.79 & 0.14 \\
\hline

\end{tabular}
\end{minipage}
\end{table*} 

\end{center}
}

\section{Results}
In the following sections we will describe the results on the
kinematics and physical conditions derived from the data. Discussion
of the emission line ratios and ionisation state of the gas will be
addressed in a future paper.

\subsection{Kinematics of the extended gaseous halo}
A key aspect of this project is to investigate the impact of the
activity on the circumnuclear material. It
is therefore necessary to establish the {\em exact} redshift of the galaxy
rest frame. The heliocentric redshift, calculated using the narrow
component of {[O III]} in the nuclear aperture (see section 4.2) is z
= 0.12351 $\pm$ 0.00008.

The [O II]$\lambda\lambda$3726,3728 emission lines are highly
extended, detected up to $\sim$ 12 kpc to the N, E, W and SW; up to
$\sim$ 15 kpc to
the NE and up to $\sim$ 20 kpc to the S.
Radial velocity profiles were constructed by fitting up to three Gaussians
(narrow, intermediate and broad components; see section 4.2) to the
doublet along the entire spatial range and are presented in Figure 3.
For simplicity, [O II] was treated as one line rather than a doublet, hence
the weighted mean of the rest wavelengths was used as the rest
wavelength of the doublet when calculating the velocity shifts.

There are no clear patterns in the spatial variations of the {[O II]}
radial velocity -- there is no steep
gradient across the nucleus and the overall shape does not resemble
the signature `s' shape expected for circular motions in a disk structure.
This, along with the 
strong asymmetry in spatial extent of the emission, suggests the
ambient gas has not settled down into a relaxed state.
Figure 3 also reveals large velocity shifts (see section 4.2), up to
$\sim$ 2000 km s$^{-1}$, between the different components fitted in
the nuclear aperture -- shifts
too large to be explained by gravitational motions and are likely to
result from jet-cloud interactions, quasar induced winds or
starburst-driven superwinds. 

We assume the narrowest component in the nucleus traces the rest frame of the galaxy
for five reasons:
\begin{enumerate}
\item The narrow component is consistent with the narrow component in
  the extended {[O II]} emission, 
the {\em only} component which can be
traced across the entire spatial range.
\item The redshift of the narrow component is consistent with the deep HI 21cm
absorption observed by \scite{mirabel89} and \scite{morganti02}
which is assumed to emanate
from the ambient, quiescent ISM in the galaxy. It is also consistent
with the redshift of the stellar absorption lines \cite{grandi77}.
\item The narrow component is by far the least kinematically disturbed
component (FWHM = 340 $\pm$ 23 km s$^{-1}$).
\item The narrow component has a small velocity amplitude
  ($\Delta_{\rmn vel}$ $\leq$ 250 km s$^{-1}$) 
  consistent with gravitational motions in low-z galaxies
  \cite{tadhunter89}. 
\item As will be shown in the next few sections, whilst different
models for the intermediate and broad components are required to
reproduce the profiles of different emission lines, the narrow components {\it
remain the same} for all models and all emission lines. The narrow
component data are presented in Table 2.
\end{enumerate}

\subsection{Modelling the emission lines}

In section 4.1 we showed the narrow component of [O II] traces the
rest frame of the galaxy. Here we model all strong emission lines in the
nucleus to search for outflows.

Prior to the modelling of the emission line profiles in the nuclear
regions a continuum model, comprising an elliptical galaxy template, a
power-law and a nebular continuum, was subtracted from the data.
The emission lines were modelled using the minimum number of Gaussians
required to produce a physically viable good fit. In order to model the
unusually broad emission lines in the nucleus  of PKS 1345+12, it
is essential to use three Gaussians (note by nucleus we mean
an aperture 2.16 arcsec wide centred on the continuum emission). 
Figure 4 shows the best fit model for the
[O III]$\lambda\lambda$4959,5007 lines. 

The [O III] lines were fitted using three constraints in accordance
with atomic physics: the flux ratio between [O III]$\lambda$4959
and [O III]$\lambda$5007 was set at 2.99:1 (based on  the transition
probabilities); the widths of the
corresponding components of each line were forced to be equal; and
the shifts between the corresponding components of each line were
fixed to be 48.0\AA. (Note, the fitting program used can only work
with a wavelength difference and not a ratio of wavelengths. However,
we find the incurred error is smaller than our estimated uncertainty.)
The best-fitting model comprises 3 components
for each line:
\begin{enumerate}
\item A narrow component, FWHM =  340 $\pm$ 23 km s$^{-1}$.
\item An intermediate component, FWHM = 1255 $\pm$ 12 km s$^{-1}$,
blue shifted
by 402 $\pm$ 9 km s$^{-1}$ with respect to the narrow component.
\item A broad component, FWHM = 1944 $\pm$ 65 km s$^{-1}$, blue shifted
by 1980 $\pm$ 36 km s$^{-1}$ with respect to the narrow component.
\end{enumerate} 
All three components are plotted in Figure 3. 

\begin{figure}
\centerline{\psfig{file=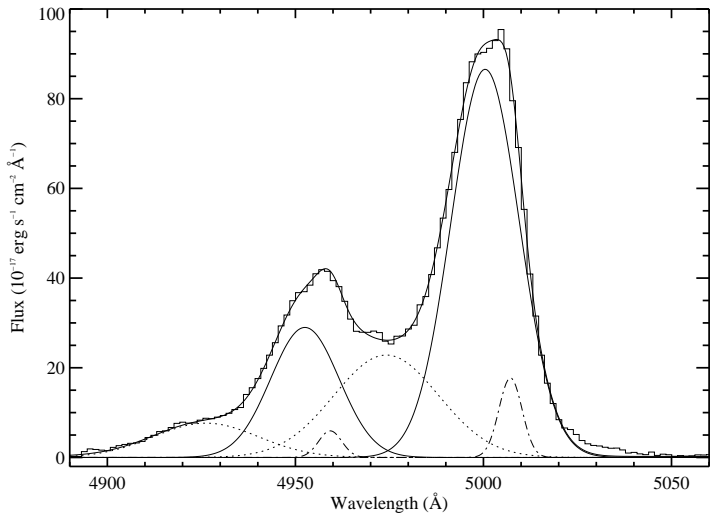,height=7.3cm}}
\caption{Model for {[O III]}$\lambda\lambda$4959,5007 in the
  nucleus. The best fitting overall
profile is overplotted (bold line) on the extracted spectrum (feint
line). All of the 6 components (3
for each line) are also plotted: the dot-dashed lines trace the narrow
components, the solid lines represent the intermediate components and the dotted
lines trace the broad components.}
\label{Figure 4.}
\end{figure}

Initially, in modelling the other emission lines,
we assumed that one model would reproduce {\em all} 
emission lines, a technique which has been successful in studies of
jet-cloud interactions in powerful radio galaxies
(e.g. \pcite{villarmartin99a}, \pcite{villarmartin99b}). Hence, we
attempted to model the other emission lines with the same three
components and the same velocity widths and shifts as {[O III]},
leaving the relative fluxes in the kinematic sub-components to vary.
We will refer to this as `the {[O III]} model' hereafter.

The {[O III]} model gives a good fit to H$\beta$ and is therefore
assumed to reproduce H$\alpha$. However, as demonstrated
in Figures 5 and 6, the {[O III]} model poorly reproduces the emission
line profiles for {[O I]}$\lambda\lambda$6300,6363 and, to a lesser
extent, {[S II]}$\lambda\lambda$6716,6731.

For {[O I]}$\lambda\lambda$6300,6363 and {[S
II]}$\lambda\lambda$6716,6731, various models were tried ranging from free
fitting (i.e. only the shift between the two {[O I]} or {[S II]} lines, 
the requirement that the corresponding components had the same width
and, in the case of {[O I]}, the intensity ratio 6300/6363 = 3.00, were
fixed) to highly constrained models to find the best
overall model. For all models, the narrow component varied little and
was consistent in velocity width and shift
with the narrow component of the {[O III]}
model. Hence, for all models for all lines, the narrow component width
and shift were constrained to be the same as that in the {[O III]}
model. Note that, it is essential to use 3 Gaussians to reproduce {\em all}
emission line profiles, irrespective of the model used.

\begin{figure*}
\begin{minipage}{130mm}
\centerline{\psfig{file=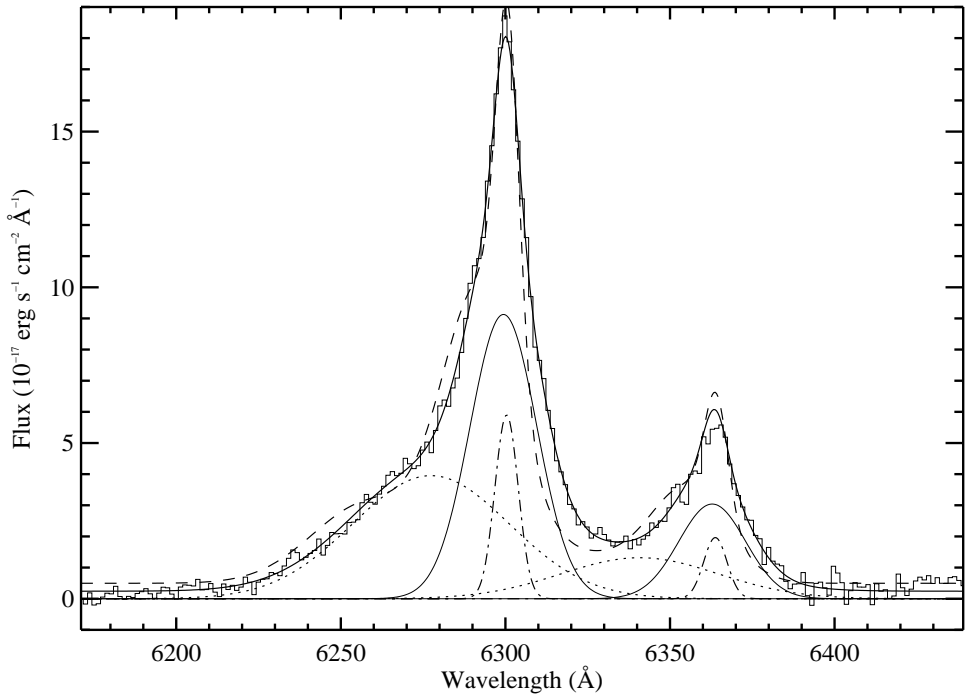,width=14cm}}
\caption{Models for {[O I]}$\lambda\lambda$6300,6363 in the nucleus. 
The bold line represents the best fitting model for {[O I]} (the {[O I]}
model) and the three components for each line are plotted (see Figure
4 for key). The dashed line represents the {[O III]} model for {[O I]}
(see section 4.2).}
\label{Figure 5.}
\end{minipage}
\end{figure*}

\begin{figure*}
\begin{minipage}{130mm}
\centerline{\psfig{file=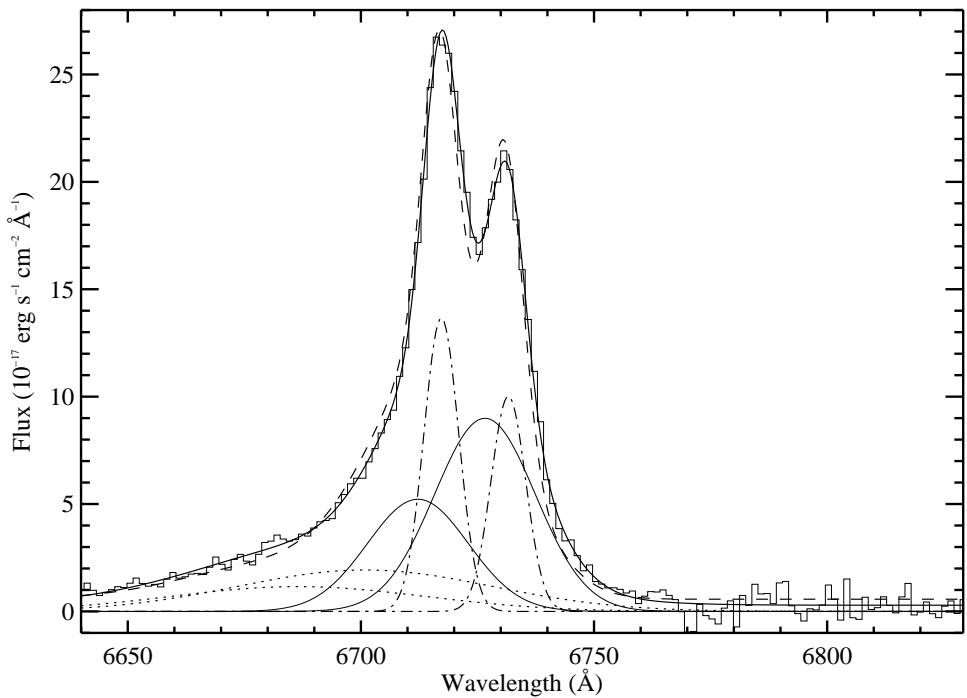,width=14cm}}
\caption{Models for {[S II]}$\lambda\lambda$6716,6731 in the nucleus. 
The bold line represents the best fitting model for {[S II]} (the {[S II]}
model) and the three components for each line are plotted (see Figure
4 for key). The dashed line represents the {[O III]} model for {[S II]}
(see section 4.2).}
\label{Figure 6.}
\end{minipage}
\end{figure*}

Figure 5 shows the best fit model for {[O I]}$\lambda\lambda$6300,6363 (the
{[O I]} model). Apart from
constraints for the narrow component, only the shift between the two
lines (63.5 \AA) and the intensity ratio ($\lambda$6300/$\lambda$6363 = 3.00),
known from atomic physics, were constrained.
The {[O III]} model for {[O I]} is also overplotted and
clearly does not reproduce the emission line profile well.

Our model for {[O I]} assumes that no emission from 
{[S III]}$\lambda$6312 and {[Fe
X]}$\lambda$6376 is significant. Models for {[O I]} 
including {[S III]}$\lambda$6312 
and {[Fe X]}$\lambda$6376
were tried with the {[O III]} model for {[O I]}. Whilst the overall line
profile was modelled well, the contribution of {[S III]}$\lambda$6312
and {[Fe X]}$\lambda$6376 became unreasonably large and the model was
discounted. 

Similarly, Figure 6 shows the best fit model for the density diagnostic {[S
II]}$\lambda\lambda$6716,6731. Again, the {[O III]} model does not
provide a good fit, but the discrepancy is not as extreme as for \oxyonew.

Due to the complex, broad emission lines in the nucleus and
uncertainties in the continuum subtraction, it was
difficult to fit the fainter lines in the nucleus and so the results
are omitted from this paper. 

\subsection{Electron densities}
\begin{figure*}
\begin{minipage}{175mm}
\centerline{\psfig{file=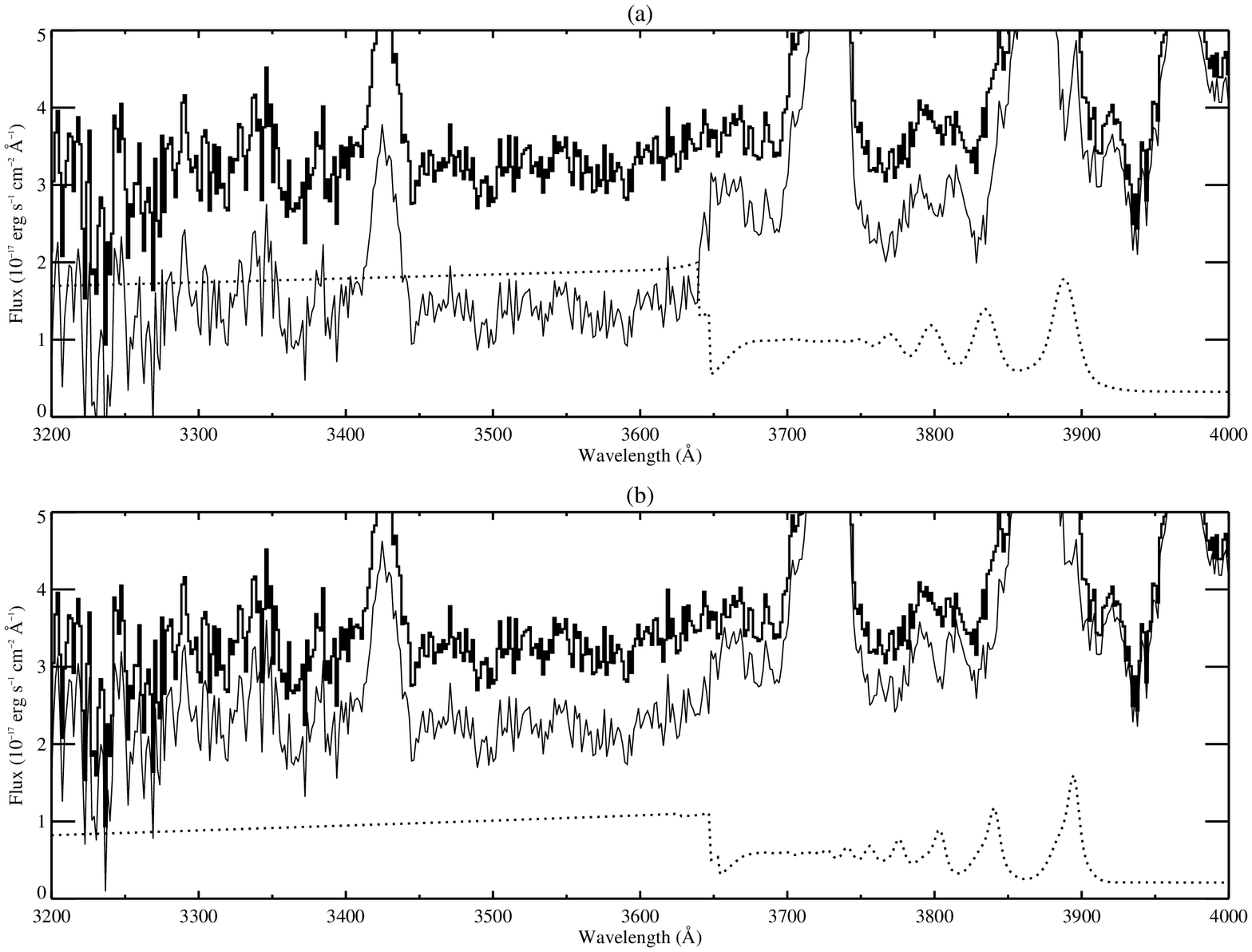,angle=90.,width=18cm}}
\caption{The nebular continuum. a) The bold line traces a section of 
the original nuclear spectrum of PKS 1345+12 in the blue. The dotted
line represents the total nebular continuum (pseudo + theoretical case
B continua for all three components) created using the measured
H$\beta$ flux (i.e. assuming zero reddening). 
The feint line traces the nebular-continuum subtracted
spectrum. There is clearly a sharp discontinuity at $\sim$ 3650 \AA~
in the nebular continuum subtracted spectrum corresponding to the jump
in the nebular continuum at the same point. This indicates that the
measured values of H$\beta$ are highly reddened resulting in an
incorrectly calculated nebular continuum. b) The same key as for a)
but the nebular continuum subtracted was generated using the reddening
corrected H$\beta$ fluxes from the E(B-V) values calculated from the
H$\alpha$/H$\beta$ ratios and then re-reddened accordingly (see section
4.4 and Table 3). This confirms the E(B-V) values for reddening are of
the correct order as there is no longer a sharp discontinuity at
$\sim$ 3650 \AA.
}
\label{Figure 7.}
\end{minipage}
\end{figure*}

\begin{center}
\begin{table*}
\begin{minipage}{115mm}
\caption{Reddening values. Columns are: $(a)$ Quantity, where: H$\alpha$/H$\beta$ is the flux 
ratio between the measured values of H$\alpha$ and H$\beta$;
E(B-V)H$\alpha$/H$\beta$ is the E(B-V) value calculated from H$\alpha$/H$\beta$ and the standard interstellar extinction curve
\protect\cite{seaton79};
E(B-V)Pa$\alpha$/H$\beta$ is the E(B-V) value calculated from
the ratio between
the value of Pa$\alpha$ quoted by \protect\scite{veilleux97} and the measured H$\beta$ flux; 
F$_{\rm H\beta}$ is the reddening corrected (assuming
the E(B-V) value calculated using the H$\alpha$/H$\beta$ ratio) H$\beta$ line flux 
(10$^{-15}$ erg s$^{-1}$ cm$^{-2}$); L$_{\rm H\beta}$ is the rest frame H$\beta$ luminosity 
(10$^{40}$ erg s$^{-1}$) and M$_{gas}$ is the mass of line emitting gas in solar masses; $(b)$ value of
particular quantity for the narrow component; $(c)$ error for narrow component; column pairs
$(d)$ and $(e)$ and $(f)$ and $(g)$ are as for $(b)$ and $(c)$. *: using quoted Pa$\alpha$ flux from 
\protect\scite{veilleux97}.
}
\label{Table 3}
\begin{tabular}{lccccrc}\\ \hline 
& narrow & $\Delta$ &intermediate&$\Delta$ & \multicolumn{1}{c}{broad} &$\Delta$\\
\multicolumn{1}{c}{$(a)$} & $(b)$ & $(c)$ & $(d)$ & $(e)$ & \multicolumn{1}{c}{$(f)$} & $(g)$  \\ \hline \hline
H$\alpha$/H$\beta$          & 3.32 & 0.33 & 5.26 & 0.28 & 18.81 & 4.74\\
E(B-V)(H$\alpha$/H$\beta$)                      & 0.06 & 0.05 & 0.42 & 0.10 & 1.44  & 0.50\\
E(B-V)(Pa$\alpha$/H$\beta$)          & - & - & -& -&  1.80(*) & 0.40 \\
F$_{\rm H\beta}$ (10$^{-15}$ erg s$^{-1}$ cm$^{-2}$)
                            & 0.70 & 0.06 & 8.83 & 0.31 & 42.66 & 0.95 \\
L$_{\rm H\beta}$ (10$^{40}$ erg s$^{-1}$)  
                            & 0.44 & 0.04 & 5.50 & 0.20 & 26.59 & 0.59 \\
M$_{gas}$ (10$^{5}$ M$_{\odot}$) 
                            & 2.61 & 0.30 & 0.92 & 0.03 & 5.64 & 1.22 \\
M$_{gas total}$ & \multicolumn{6}{c}{6.54 $\pm$ 1.22
  $\times$ 10$^{5}$ M$_{\odot}$} \\
\hline
\end{tabular}
\end{minipage}
\end{table*}
\end{center}

During the early stages of radio source evolution, the nuclear regions
of radio sources are likely to harbour a rich ISM, possibly left over
from a recent merger. Here we investigate the
density of the line emitting gas using 
{[S II]} 6716/6731, the density diagnostic ratio.

In section 4.2 and Figure 6 we showed that the {[O III]} model fit to the
{[S II]}$\lambda\lambda$6716,6731 doublet is not wholly satisfactory.
The initial free fit for {[S II]} (i.e. only the narrow component width
and position was constrained) produced a good fit to the overall line
profile, but the ratio  was
outside the range of allowed values for both the intermediate and
broad components ({[S II]} 6716/6731 $<$ 0.44 -- the ratio
at the high
density limit). For the next attempt, the same constraints were
applied with the addition of the ratios of the broad and intermediate
components set at the high density limit (6716/6731
= 0.44). This produced a good overall fit for the doublet which was
physically viable (the narrow component ratio was well within the
allowed range).

For subsequent fits, the {[S II]} 6716/6731 intensity ratio for 
the broad and
intermediate components was slowly increased (i.e. the density
decreased) until the model broke down and no longer fitted the blend
well. Hence we estimate upper limits for the intensity ratio {[S
II]} 6716/6731 of 0.57 and 0.60 for the
intermediate and broad components respectively. Reassuringly, the
density estimates obtained by fitting the \sulphurtwo~doublet with the
\oxythr~model were similar to those obtained using the \sulphurtwo~model.

The density was calculated using the IRAF program TEMDEN which is
based on the five-level atom calculator developed by
\scite{derobertis87}, assuming an electron temperature of 10 000 K. 
Hence, upper limits for the intensity ratio {[S
    II]} 6716/6731 correspond to lower limits in density and we
calculate densities of $n_{e}$ $>$ 5300 cm$^{-3}$ and $n_{e}$ $>$
4200 cm$^{-3}$ for the intermediate and broad components
respectively. The {[S II]} 6716/6731 ratio for the narrow component is
consistent with the low density limit and  we estimate an upper
limit of n$_{e}$ $<$ 150 cm$^{-3}$ for the narrow component.

\subsection{Reddening}
In section 4.3 we showed that the nuclear regions of PKS 1345+12
harbour a dense ISM (n$_{e}$ $>$ 4000 cm$^{-3}$). 
Observationally, large amounts of gas and
dust will be detected as reddening of the optical spectrum.

We have investigated reddening in PKS 1345+12 using three independent
techniques: the H$\alpha$/H$\beta$ ratio, the Pa$\alpha$/H$\beta$ ratio
and nebular continuum subtraction, assuming a simple foreground screen
model for interstellar dust.

\subsubsection{Balmer line ratios}
The degree of reddening can be estimated using the Balmer line ratio
H$\alpha$/H$\beta$. Note that the H$\alpha$/{[N
II]}$\lambda\lambda$6548,6583 blend in the nuclear spectrum of PKS
1345+12 is highly complex -- all three lines are broad, have blue
asymmetries and require three components each. Initially,
both the H$\alpha$ and {[N II]} lines were fitted
using the {[O III]} model. The {[O III]} model was used for H$\alpha$
since H$\beta$ is well modelled by it. For {[N II]}, various models
were tried and the {[O III]} model produced the best overall fit for
the blend. Using this {[O III]} model we measured 
H$\alpha$/H$\beta$ ratios of 3.32 $\pm$ 0.33, 5.26 $\pm$
0.28 and 18.81 $\pm$ 4.74 leading to E(B-V) values
 of 0.06 $\pm$ 0.05,
0.42 $\pm$ 0.10 and 1.44 $\pm$ 0.50 for the narrow, intermediate and broad
components respectively (see Table 3; \pcite{seaton79} extinction law assumed).

Given the complexity of the blend, 
we also re-modelled the lines to find the
lowest possible flux for H$\alpha$, hence a lower limit for the ratio
H$\alpha$/H$\beta$ and the E(B-V) value. We used the {[O III]} model
for H$\alpha$ but for {[N II]} only constrained the narrow components
and for the intermediate and broad components, the flux ratio ({[N
II]} 6583/6548 = 3.00) and
the shift between the lines (35.27 \AA)
and forced the velocity widths of the corresponding components to be
equal. This technique gave lower limits for H$\alpha$/H$\beta$ of
H$\alpha$/H$\beta$ $>$ 3.91 and H$\alpha$/H$\beta$ $>$ 9.81 
for the intermediate and broad components respectively
corresponding to lower limits for E(B-V) of E(B-V) $>$ 0.19 and E(B-V)
$>$ 0.92 for the
intermediate and broad components respectively.

\subsubsection{Comparison of the optical and infra-red}
\scite{veilleux97} detected two components to Pa$\alpha$: broad (FWHM $\sim$ 2600 km s$^{-1}$) 
and narrow (FWHM $\sim$ 900 km s$^{-1}$), and they identified the broad component
as broad line region (BLR) emission from a lightly-extinguished
quasar nucleus. Although this interpretation is consistent with the detection
of a point source in high-resolution near-IR observations \cite{evans99},
there must be some doubt about it given the detection of broad components
to the forbidden lines at optical wavelengths. Unfortunately, it is difficult to 
establish the link between the various optical and infrared kinematic
components with greater certainty, because \scite{veilleux97} neither
publish the line shifts, nor fit an intermediate width component. However, by
assuming that {\em all} the flux in the broad Pa$\alpha$ is associated
with the same component that emits the broad H$\beta$ detected in our
optical spectrum and by varying this flux by a factor of 2 to account for 
possible slit-loss differences between the observations, 
we obtain an upper limit for reddening of the optical 
broad component of E(B-V) $<$ 2.0 -- consistent with the other estimates
obtained above.

\subsubsection{The nebular continuum}
Another indication of large reddening in the nuclear spectrum of PKS
1345+12 comes from consideration of the nebular continuum, 
calculated as part of the continuum modelling process prior to
measurement of the emission lines.
The nebular continuum for each kinematic
component was created using the technique described
in \scite{dickson95}. The nebular continuum strength is not
particularly sensitive to temperature but we have assumed an electron
temperature, T$_e$ = 10 000 K.

Initially, we assumed zero reddening for all components. 
Figure 7$a)$ shows the original spectrum, the 
zero reddening nebular continuum
and the nebular continuum subtracted spectrum in the nucleus of PKS
1345+12 in the blue. 
After subtracting the nebular
continuum, a clear discontinuity is left at the position of the Balmer
edge at $\sim$ 3645\AA~which is
unphysical. This indicates that the nebular continuum has been
over-estimated in the zero reddening case
and that reddening is likely to be significant in the nucleus
of PKS 1345+12.

Using the E(B-V) values calculated from the measured
H$\alpha$/H$\beta$ ratios (see section 4.4.1) and the standard interstellar extinction
curve \cite{seaton79}, we corrected the H$\beta$ fluxes for reddening
and calculated the reddened nebular continuum spectra.

Figure 7$b)$ shows the original spectrum, 
the new nebular continuum (assuming E(B-V) values of 0.06, 0.42 and
1.44 for the narrow, intermediate and broad components respectively)
and the nebular continuum subtracted spectrum in the nucleus of PKS
1345+12. 
Clearly, when the reddened nebular continuum is subtracted,
there is no longer a discontinuity. This is further evidence for large
reddening in the nucleus of PKS 1345+12

To further investigate the degree of reddening in the broad and intermediate
components (the amount of reddening in the narrow component is
negligible), this process was repeated for various E(B-V) values
for the intermediate and broad components.
The intermediate component contributes most of the flux in the nebular
continuum and we estimate a lower
limit of E(B-V) $>$ 0.3 for this component 
with some confidence -- at this level of
reddening, the discontinuity in the nebular continuum subtracted
spectrum disappears. 

The variation of the
broad component had negligible effect on the resultant nebular
continuum. Hence, the lower limit of E(B-V) $>$ 0.92 for the broad
component comes from the calculated Balmer line ratios.

To summarise, the
nuclear regions of PKS 1345+12 harbour a rich ISM and are subject
to large reddening. Each of the three independent techniques for
estimating reddening give consistent results and it is clear there is
a strong correlation between the broadness of component and the degree
of reddening, in the sense that the broadest components are the most
highly reddened.

\subsection{Estimating the gas mass}
We have seen that the nuclear regions of PKS 1345+12 harbour a rich,
dense ISM, but
is there sufficient mass to confine and frustrate the radio source?
Emission line luminosities are related to the mass of line emitting
gas by:
\[
\rmn{M_{gas}} = \rmn{m_p} \frac{\rmn{L}(\rmn{H}\beta)}{\rmn{N_e}
\alpha_{\rmn{H}\beta}^{eff} h\nu_{\rmn{H}\beta}}
\]
where N$_e$ is the electron density
(cm$^{-3}$); m$_p$ is the mass of the proton (kg);
L(H$\beta$) is the luminosity of the H$\beta$ line (erg s$^{-1}$);
$\alpha_{H\beta}^{eff}$ is the effective recombination coefficient for
H$\beta$ (cm$^3$ s$^{-1}$) and h$\nu_{H\beta}$ is the energy of an
H$\beta$ photon (erg).

In section 4.3 we calculated densities of n$_{e}$ $<$ 150 cm$^{-3}$, n$_{e}$ $>$ 5300
cm$^{-3}$ and n$_{e}$ $>$ 4200 cm$^{-3}$ in the regions emitting the narrow,
intermediate and broad components respectively. Using the presented
equation, lower limits in density correspond to upper limits in
mass. Hence, we estimate limits on the mass of line emitting gas of
M$_{gas}$ $>$ 2.61 $\times$ 10$^{5}$ M$_{\odot}$, M$_{gas}$ $<$ 0.92 $\times$ 10$^{5}$
M$_{\odot}$ and  M$_{gas}$ $<$ 5.64 $\times$ 10$^{5}$ M$_{\odot}$ in the regions
emitting the narrow, intermediate and broad components respectively
(see Table 3). This leads to an upper limit of order 10$^{6}$
M$_{\odot}$ for the total mass of line emitting gas in the
kinematically disturbed emission line components.

\section{Discussion}
\begin{figure*}
\begin{minipage}{85mm}
\centerline{\psfig{file=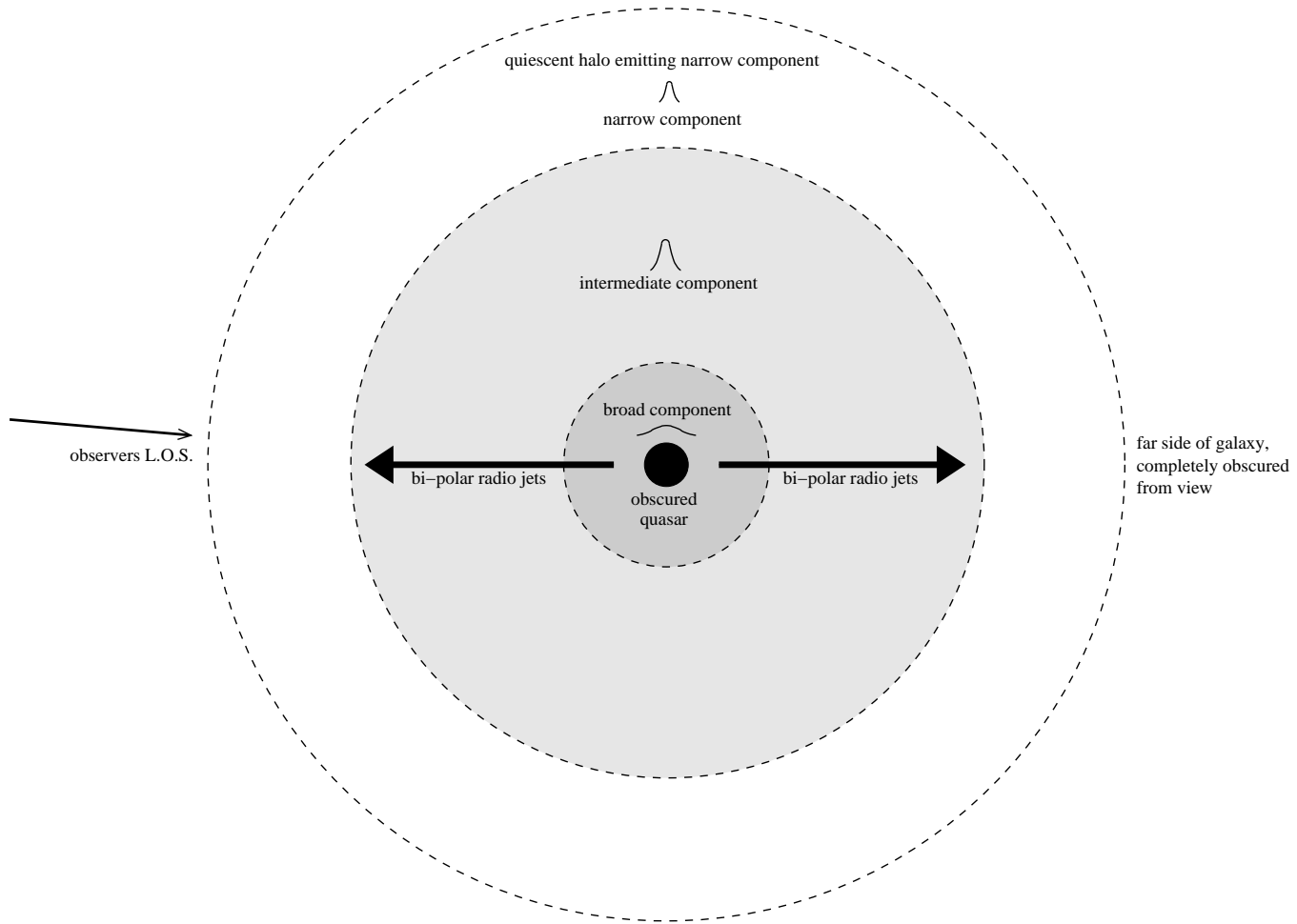,width=18cm}}
\caption{Schematic diagram of the model for PKS 1345+12.
}
\label{Figure 8.}
\end{minipage}
\end{figure*}

\subsection{A model for PKS 1345+12}
The broad, multiple-component emission lines in PKS 1345+12 indicate
that line emission emanates from {\em at least} three kinematically distinct
regions in the NLR along the line of sight. To determine what these
components represent, it is necessary to establish the exact redshift
of the galaxy rest frame.

In section 4.1, we argued that the narrow component represents the
ambient, quiescent ISM in the galaxy halo. We are confident of this
result -- the narrow component is the only component consistently
found in all emission lines, irrespective of model required or spatial
position; kinematically, it is the least disturbed component and it is
consistent with the deep, narrow HI 21cm absorption
trough (\pcite{mirabel89}, \pcite{morganti02}) 
usually associated with the ambient, quiescent gas halo
surrounding radio sources.
Hence, the blue asymmetries in the emission line
profiles represent material moving towards us, blue shifted with
respect to the galaxy. 

In sections 4.3 and 4.4 we showed that the intermediate and broad
components originate from regions that are dense (n$_e$ $>$ 4000 cm$^{-3}$)
and dusty (E(B-V) $>$ 0.30 and E(B-V) $>$ 0.92 for the intermediate and
broad components respectively). 
It is unlikely, therefore, that we would see
emission from material on the far side of the nucleus -- the further
away from the observer in PKS 1345+12 along the line of sight, the
more extinction the emission will suffer. The blue shifted
material in PKS 1345+12 is therefore likely to 
be material in outflow on the side of the
nucleus closest to the observer. 

Indeed, similar conclusions have been
made about the compact, flat-spectrum radio source PKS 1549-79 by
\scite{tadhunter01}. The optical spectrum of PKS 1549-79 shows many
characteristics similar to PKS 1345+12, namely blueshifted broad 
emission lines. Moreover,
a large extinction for this source is suggested by the fact that,
whilst the flat radio
spectrum indicates the presence of a QSO, there is no evidence at
optical wavelengths
for the
broad permitted lines and non-stellar continuum characteristic of
QSOs. We will now build on the simple model proposed for PKS 1549-79
to account for more complex characteristics observed in PKS 1345+12.

In sections 4.2 and 4.3 we showed that three Gaussians were 
required to fit all the emission lines, irrespective of the model required
to reproduce the profile. The models vary in both velocity width and
shift of the intermediate and broad components with respect to the
narrow component. We interpret this as three kinematically and physically distinct
regions in PKS 1345+12, each responsible for the emission of one of
the components (i.e. narrow, intermediate or broad) although gradients
in ionisation/density, coupled with gradients in velocity 
must exist across these regions to produce different broad or
intermediate components (see Figure 8). 

We can determine the locations of the three kinematically distinct
regions using the reddening results. In section 4.4 we showed that
whilst there is significant reddening in the nuclear spectrum of PKS
1345+12, the amount of reddening in the three components differs
substantially. The narrow component has negligble reddening and must
therefore originate from the outer regions of the galaxy. This is
consistent with other evidence suggesting the narrow component
originates from the galaxy halo (see section 4.1 and
above). Similarly, the intermediate and broad components are subject
to significantly more reddening with the degree of reddening correlating with
the broadness of the component. The broadest component is therefore
likely to 
originate from the regions closest to the galactic centre and the
intermediate component slightly further out but within the narrow
component emitting region.

We now propose an evolutionary scenario for PKS 1345+12:
\begin{enumerate}
\item PKS 1345+12 underwent a merger event in its recent history. This
  merger consisted of two galaxies (PKS 1345+12 has a double nucleus),
 at least one of which was gas rich. The merger event started so
  recently it is still ongoing, indicated by the presence of two nuclei
(e.g. \pcite{gilmore86}; \pcite{heckman86}; \pcite{odea00}).
This is consistent with the idea
  that many radio galaxies appear to have been triggered by merger events.
\item The merger event provided large amounts of gas and dust to the
  regions surrounding the nucleus, obscuring it from view. 
\item The current activity was triggered in the the western nucleus,
  the nucleus coincident with the radio source.
Note, we use the term `current activity' because, whilst we can say
  with confidence that the radio source is young (see section 5.2), 
we cannot rule out
  the possibility of previous activity phases which may be responsible
  for the extended, diffuse halo emission.
\item While the active nucleus is obscured, the outer parts (halo etc) of the
  galaxy start to
settle down. The halo is responsible for the
emission of the narrow component and the deepest HI absorption trough
(\pcite{mirabel89}; \pcite{morganti02}). Hence, the velocity width of
 the narrowest component is consistent with gravitational motions.
PKS 1345+12 is a GPS source and is therefore observed relatively
  shortly after the triggering of the radio activity.
With time, the radio source will expand, first becoming
a larger compact steep spectrum radio source (CSS) before becoming extended.
\item As the radio jets expand, material is shocked, accelerated and 
pushed aside as the
jets hollow out cavities on either side of the quasar (bi-polar). 
As well as jet-cloud interactions, quasar- or starburst-induced winds
are also likely to be important at this stage. Indeed, outflow
components with velocity shifts similar to those measured in PKS
1345+12 have also been detected in ULIRGs without strong radio
emission \cite{wilman99}. Therefore, the dominant driving mechanism
for the outflows is currently unknown.
\item Eventually, the radio jets will tunnel through the cocoon and
expand freely to evolve into a large scale, extended radio galaxy.\\
\end{enumerate}

\subsection{PKS 1345+12 - young radio source}
A crucial question regarding compact radio sources involves the amount
of material in their circumnuclear regions. In the evolutionary
scenario outlined above, we have assumed that PKS 1345+12 is a young
radio galaxy. It is clear that the circumnuclear ISM in PKS 1345+12 
is both dense and dusty, but is there sufficient mass to confine and 
frustrate the radio source?

In section 4.5, we estimated the mass of line emitting gas using the
reddening corrected H$\beta$ luminosities (see Table 3). We estimated
an upper limit for the total mass of line emitting gas (i.e. for the
kinematically disturbed components combined) of $<$ 10$^{6}$ M$_{\odot}$.

Several simulations have been run varying the distribution and
density of the insterstellar medium to determine the mass of gas
required to frustrate a radio source. Hydrodynamical simulations
carried out by \scite{deyoung93} showed that for low to intermediate
power radio sources (10$^{43}$ - 10$^{44}$ erg s$^{-1}$) in both
a homogeneous and a two-phase ISM, 
gas masses of order 10$^{11}$ M$_{\odot}$
were sufficient to confine the radio source for its entire
lifetime. For the high power radio sources (10$^{45}$ erg s$^{-1}$),
the masses required would be too large to be credible.
Analytical models of jet interactions by
Carvalho (1994,1998) suggest confinement is possible with much lower
gas masses, of order 10$^{9}$ - 10$^{10}$ M$_{\odot}$ if the ISM is 
clumpy and within a scale of hundreds of pc of the radio
source. 

At face value, these simulations, considering either a smooth medium or
 small clouds evenly distributed throughout a sphere,
 may lead us to conclude that our
estimate of M$_{gas}$ $<$ 10$^{6}$ \msun~may be too small to confine
and frustrate the radio source and that PKS 1345+12 is a young radio
source. However, if the ISM forms large dense clouds, much smaller masses
 could significantly disrupt the propagation of the jet (e.g. forming
 bubbles as in M87; \pcite{bicknell96}) which may lead to some degree of
frustration.

\section{Conclusions and future work}
PKS 1345+12 is most likely a young radio source observed sometime
after the start of the merger event which triggered the nuclear
activity. We observe complex, multiple component emission lines in the
nucleus which, due to large reddening in the nucleus, 
we interpret as material in outflow. Line emission
originates from three kinematically distinct regions over which exist
gradients in velocity, acceleration and possibly ionisation potential.
The major outstanding issues for PKS 1345+12 concern the acceleration
and ionisation mechanisms for the gas. In particular, whether the
outflows are driven by the jets, AGN driven winds or starburst driven
super-winds. In future, it should be possible to address these issues
using high resolution imaging observations of the outflow regions in
combination with detailed studies of the emission line ratios derived
from our spectra.

Overall, our results provide evidence that the activity has a major
impact on the circum-nuclear gas in the early stages of radio source
evolution. 

\section*{\sc Acknowledgements}
JH acknowledges a PPARC PhD studentship. We thank Geoff Bicknell for
very useful discussions regarding the expansion of radio jets in dense
environments. We also thank the referee, Dr. Ignas Snellen, 
for useful comments on the paper.
The William Herschel Telescope is operated on the
island of La Palma by the Isaac Newton Group in the Spanish
Observatorio del Roque de los Muchachos of the Instituto de
Astrofisica de Canarias. This research has
made use of the NASA/IPAC Extragalactic Database (NED) which is
operated by the Jet Propulsion Laboratory, California Institute of
Technology, under contract with the National Aeronautics and 
Space Administration. Figures 2 to 7 inclusive were created in
IDL. The preliminary results of this study were published in PASA,
Volume 20.

\bibliographystyle{mnras}
\bibliography{abbrev,Holt_etal_refs}

\end{document}